\pdfoutput=1

\documentclass[sigconf]{acmart}

\usepackage[utf8]{inputenc}
\usepackage{booktabs} 
\usepackage{multirow}
\usepackage{graphicx}
\usepackage{subfigure}
\usepackage{tabularx}
\usepackage[flushleft]{threeparttable}
\usepackage{multirow}
\usepackage{natbib}

\usepackage{algorithm}
\usepackage{adjustbox}
\usepackage[noend]{algpseudocode}

\usepackage{xcolor}

\renewcommand\footnotetextcopyrightpermission[1]{} 
\begin{document}
\title{Is it a work or leisure travel? Applying text classification to identify work-related travel on social networks}

\author{Lucas Félix}
\affiliation{%
    \institution{Universidade Federal de Minas Gerais}
    \country{Brasil}
}
\email{lucas.felix@dcc.ufmg.br}

\author{Washington Cunha}
\affiliation{%
    \institution{Universidade Federal de Minas Gerais}
    \country{Brasil}
}
\email{washingtoncunha@dcc.ufmg.br}

\author{Jussara Almeida}
\affiliation{%
    \institution{Universidade Federal de Minas Gerais}
    \country{Brasil}
}
\email{jussara@dcc.ufmg.br}

\fancyhead{}

\renewcommand{\shortauthors}{L. Félix et al.}



\copyrightyear{2023}
\acmYear{2023}
\setcopyright{othergov}

\maketitle

\section{Introduction}
    
    Through the internet, users gather in forums, communities, travel blogs, and \textit{Social Networks} (\textbf{SNs}). Indeed, these platforms are useful tools considering that they made it possible to seek \textit{Points of Interest} (\textbf{POI}), provide and receive information about places, and share experiences from a single point of view. Studies show that $80\%$ of American travelers use \textit{SNs} while traveling, and more than half of this percentage share their journey information with their contacts~\cite{yoo2016exploring}. This study also showed 'how' and 'why' \textit{SNs} are the tool most adopted by usual travelers.\looseness=-1

    Between the platforms available on the web Location-Based SN (\textbf{LBSN}) and Travel SN (\textbf{TSN}) are more focused on tourism and have been extensively used in the literature in tourism works~\cite{khatibi2018improving}. The LBSNs' main purpose (e.g., \textit{Yelp} and \textit{Foursquare}) focus on the user's current location sharing their footsteps. On the other hand, the \textit{TSNs} (e.g., \textit{TripAdvisor}) are responsible for joining in the same platform several pieces of information, such as locations, accommodations, transport, food, attractions, and services~\cite{yoo2016exploring}, enabling their users to describe better the full trip experience. Furthermore, \textit{TSNs} allows users to plan trips, look at different perspectives on a place, and get recommendations about users with the same taste~\cite{kim2009triptip}, while comparing prices.\looseness=-1

    Although TSN and LBSN platforms concentrate most of the information needed to plan a full trip experience to a specific place (or set of places), users still encounter difficulty with the huge amount of available data, precluding to distinguish which option (or set of options) is the best~\cite{pang2008opinion}. Therefore, \textit{Recommender Systems} (\textbf{RS}) solutions are usually applied to tackle the information overload problem, focusing on suggesting POIs based on the user history. RS approaches reduce the number of options for the users and enable them to choose between a smaller (and possibly better) option set.

    Many widely studied aspects, such as physical constraints, social and temporal influences, make the POI recommendation scenario harder than the classic one. Consequently, the literature has naturally adopted contextual information to leverage recommendations quality. Indeed, additional data allows for more accurate recommendations than traditional methods (e.g., Collaborative Filtering)~\citep{wang2013location}.\looseness=-1

    One additional piece of information that can be used for tourism recommendations is the user \textbf{travel purpose}. In the literature, this has not been properly exploited due to the fact that this is not usually available in datasets and, when available, is not usually filled by the users. On the other hand, this can be used to properly characterize tourists' actual behavior while traveling ~\citep{dietz2019characterisation}, and providing, in this way, a more suited recommendation for these users. \looseness=-1

    Accordingly, in this work, we propose a model to predict whether a trip is leisure or work-related. We aim to enable authors to characterize better travels made by users and perform better recommendations in specific scenarios. To accomplish so, we compare different state-of-the-art \textbf{Automatic Text Classification (ATC)} models, namely BERT~\cite{devlin2018bert}, RoBERTa~\cite{liu2019roberta}, and BART~\cite{lewis2019bart}.\looseness=-1

\section{Proposed Strategy}

    \subsection{Data Collection and Pre-Processing}

          From the datasets available in the literature, only a few present essential features for real-world evaluation, such as the POI working hours, availability, and cost. In this scenario, data from TSN, like TripAdvisor, are more suited for such analyzes. Besides containing more information, TripAdvisor is currently the most popular travel website~\cite{khatibi2018improving}, with about $390$ million monthly unique visitors. Therefore, we adopted TripAdvisor review datasets to train an ATC model capable of distinguishing leisure and work-related travels.
          
          To accomplish so, we extracted TripAdvisors' data through a web crawler responsible for automatically browsing the website and collecting all users' available content and POIs. The collected data was firstly in an unstructured format (e.g., HTML). Then, it was parsed to retrieve the content within each page, pre-processed, and stored in semi-structured format (e.g., CSV) data.

          Our collection was initially focused on users from five different touristic cities worldwide and their complete visit history (including other cities): Tiradentes and Ouro Preto (Brazil), San Gimignano, Cannes, and Ibiza \footnote{TripAdvisor Complete Dataset}. Since our dataset possessed users' complete history, containing the visits to numerous different cities, in this work, we focused on the English reviews available \footnote{TripAdvisor English} that are associated with the trip label (e.g. family, romantic, friends, work-related, alone) \footnote{TripAdvisor English w/ classes}. We focused the analysis on English reviews due to the fact that the text classification algorithms employed in our methodology demonstrate better performance with English rather than in other languages~\cite{wang2019cross}. To identify the English reviews on the datasets used in this work, we adopted the pre-trained model proposed in ~\citep{joulin2016fasttext}.  
          
          To augment our dataset, we labeled visits occurring within the same city and time period (month and year) that lacked a label. To assign a label, we utilized the available classification from another POI visited by the same user in the same city and time period.

          Considering our aim is to define leisure and work-related travels to characterize travelers further initially, we modeled our problem as a binary classification problem by aggregating non-work classes as leisure, leaving only two classes. Following the aggregation process, it was observed that $87.67\%$ of the instances consisted of leisure-related reviews, whereas $12.33\%$ were categorized as work-related reviews. This indicates an imbalanced dataset context, where most instances pertain to leisure-related reviews.
         Table~\ref{t:ta_descriptive} presents a summary of TripAdvisor data used in this work.

        \vspace{-.3cm}
          \begin{table}[h]
            \tiny
              \centering
              \begin{tabular}{|c|c|}
                    \hline
                   \textbf{Dataset}                & \textbf{\# Instances}  \\ \hline
                    TripAdvisor Complete           &  $11,443,663$           \\ \hline
                    TripAdvisor English            &  $2,434,252$           \\ \hline
                    TripAdvisor English w/ classes &  $639,997$             \\ \hline
              \end{tabular}
              \caption{TripAdvisors' data summary}
              \label{t:ta_descriptive}
          \end{table}
        \vspace{-1cm}

    \subsection{Text Classification}

        In the ATC block of our modeling, we aim to classify the users reviews, identifying if their travels were leisure or work-related. To accomplish so, we compare three different neural approaches that are the state-of-the-art on text classification task, achieving the best results in the benchmarks used by researchers~\cite{de2023class,cunha2023tpdr}, namely we use BERT~\citep{devlin2018bert}, RoBERTa~\citep{liu2019roberta} and BART~\citep{lewis2019bart}.

        To compare the effectiveness of each approach, we follow the procedure defined in~\citep{cunha2022comparative}. We evaluate the effectiveness of the proposals using Macro and Micro Averaged F1. The experiments were executed using a $5$ fold cross-validation procedure. To compare the cross-validation results, we evaluate the statistical significance using a paired t-test with $95\%$ confidence with
        Bonferroni correction to account for the multiple tests~\cite{mendes2020keep,cunha2023effective}.

        As mentioned earlier, our dataset exhibits a significant skew, making it challenging to obtain accurate predictions for both classes. To address this issue, we implemented a class balancing procedure during the training phase for each fold, in which we randomly chose $N_{mino}$ instances from the majority class, ensuring equilibrium with the $N_{mino}$ instances derived from the minority class.
        During the testing phase, on the other hand, we maintained the overall distribution of the datasets to ensure a realistic representation of the actual data.\looseness=-1

\section{Experimental Evaluation}\label{sec:experimental}


    In this section, we present the results achieved by our proposal. First, we present and discuss some instances used to train the models. Then, we present the metrics for each model. 

    \subsection{Dataset}

    In Table~\ref{t:examples}, we present some examples of each class present in our dataset. In these examples, it is easy to distinguish each class, especially in the second case, due to the fact that the author states that it is work-related travel. Nevertheless, not all work-related instances can be easily distinguished, as in this example. Illustrating that, we have the following example, which is work-related travel:
    \textit`{"I don't really understand much of the exhibits, but it can be an eye-opener. The free exhibits took us slightly less than 2hrs to complete. The navigation is easy. Premise is clean and spacious."}. However, in this particular example, there are no indications suggesting that this trip is work-related. Therefore, as part of our future work, we aim to incorporate additional features that can help us differentiate between leisure and work-related trips. Our intention is to leverage features that are commonly found across various datasets, thereby enabling the application of our proposed model in different scenarios.

     \begin{table}[h]
            \tiny
            \begin{tabular}{|c|c|}
            \hline
            \textbf{Class}             & \textbf{Review} \\ \hline
            \textbf{Leisure} & "What a fantastic example of the Brazilian Barroc art..."  \\ \hline
            \textbf{Work} & "..This hotel is one of my favorite I've stayed in for work travel..." \\ \hline
            \end{tabular}
            \caption{Instances used to train in each class}
            \label{t:examples}
            \vspace{-1cm}
    \end{table}

    \subsection{Classification Results}
    
    Table~\ref{t:results} results show the effectiveness of each evaluated model, presenting Macro-F1, Micro-F1, and the Confidence Interval of each algorithm. Even though the RoBERTa approach presents the best average results, all the algorithms are statistically equivalent.\looseness=-1

    \vspace{-.3cm}
    \begin{table}[h]
    \tiny
    \centering
        \begin{tabular}{|c|c|c|}
            \hline
            \textbf{Model} & \textbf{Macro-F1} & \textbf{Micro-F1} \\
            \hline
            \textbf{Bart} & 69.1(1.52)      & 80.86(1.89)  \\
            \textbf{Bert} & 67.36(1.45)     & 79.78(1.36) \\
            \textbf{RoBerta} & 70.16(1.57)  & 82.15(2.11)  \\
            \hline
        \end{tabular}
        \caption{Models metrics and CI}
        \label{t:results}
    \end{table}
    \vspace{-0.9cm}

    Based on the results, it can be inferred that the algorithms demonstrate strong predictive capabilities for both classes. The evaluation of Macro-F1 metrics supports the notion that the algorithms achieve favorable classification outcomes for both classes. However, as mentioned earlier, we acknowledge that in certain instances, relying solely on reviews may not provide sufficient discernment to accurately differentiate trip labels. Consequently, this limitation has an impact on the Macro-F1 results.

\begin{acks}
This work was supported by Conselho Nacional de Desenvolvimento Científico e Tecnológico (CNPq), Fundação de Amparo à Pesquisa do Estado de Minas Gerais (FAPEMIG), Coordenação de Aperfeiçoamento de Pessoal de Nível Superior (CAPES) and the CAPES-STIC-AMSUD 22-STIC-07 LINT project.
\end{acks}

\bibliographystyle{unsrtnat}
\bibliography{main}

\end{document}